\documentclass{article}
\usepackage{spconf,amsmath,graphicx}
\usepackage{enumitem}
\usepackage{url}
\usepackage[page]{appendix}
\usepackage{hyperref}
\usepackage{multirow}
\usepackage{booktabs}
\hypersetup{colorlinks=true}
\usepackage{color}
\usepackage{xcolor}
\usepackage{subcaption}
\usepackage{diagbox}
\usepackage[export]{adjustbox}

\newcommand{\blank}{\texttt{<blank>}}

\newcommand{\footnoteurl}[1]{\footnote{\scriptsize \url{#1}}}
\def\bbR{{\mathcal{R}}}
\def\calY{{\mathcal{Y}}}

\title{Towards Word-Level End-to-End Neural Speaker Diarization with Auxiliary Network}
\name{Yiling Huang, Weiran Wang, Guanlong Zhao, Hank Liao, Wei Xia, Quan Wang}
\address{
Google LLC, USA \\
\vspace{-5mm}
\\\texttt{\normalsize \{
\href{mailto:yilinghuang@google.com}{yilinghuang},
\href{mailto:weiranwang@google.com}{weiranwang},
\href{mailto:guanlongzhao@google.com}{guanlongzhao},
\href{mailto:hankliao@google.com}{hankliao},
\href{mailto:ericwxia@google.com}{ericwxia},
\href{mailto:quanw@google.com}{quanw}
\}@google.com}}
\begin{document}
\ninept
\maketitle
\begin{abstract}
While standard speaker diarization attempts to answer the question ``who spoken when'', most of relevant applications in reality are more interested in determining ``who spoken what''. Whether it is the conventional modularized approach or the more recent end-to-end neural diarization (EEND), an additional automatic speech recognition (ASR) model and an orchestration algorithm are required to associate the speaker labels with recognized words. In this paper, we propose Word-level End-to-End Neural Diarization (WEEND) with auxiliary network, a multi-task learning algorithm that performs end-to-end ASR and speaker diarization in the same neural architecture. That is, while speech is being recognized, speaker labels are predicted simultaneously for each recognized word. Experimental results demonstrate that WEEND outperforms the turn-based diarization baseline system on all 2-speaker short-form scenarios and has the capability to generalize to audio lengths of 5 minutes. Although 3+speaker conversations are harder, we find that with enough in-domain training data, WEEND has the potential to deliver high quality diarized text.

\end{abstract}

\section{Introduction}
\label{sec:intro}

Speaker diarization is the task of partitioning speech into homogeneous segments according to speaker identities. The traditional approach is a combination of multiple individually trained modules, including voice activity detection (VAD), speaker turn segmentation, speaker encoder and clustering. Each module has been extensively studied to improve speaker diarization, including personalized VAD~\cite{medennikov2020target}, better speaker turn detection~\cite{xia2022turn}, fine-tuning speaker encoders for specific scenarios (e.g. ECAPA-TDNN~\cite{dawalatabad2021ecapa} for short queries), and various clustering algorithms~\cite{wang2017speaker,dimitriadis2017developing,garcia2017speaker}.
More recently, the research community has been exploring supervised end-to-end approaches including UIS-RNN~\cite{zhang2019fully}, DNC~\cite{li2019discriminative}, frame-level end-to-end neural diarization (EEND)~\cite{fujita2019end}, and its other variants~\cite{horiguchi2020end,fujita2020neural,xue2021online,han2021bw,kinoshita2021integrating}. Other methods are described and discussed in literature reviews and tutorials~\cite{park2021review,zhang2022odysseytutorial}.

Most of the above mentioned speaker diarization systems output timestamped segment-level speaker labels (i.e. ``who spoke when''), which are usually not useful by themselves. For most real-world applications, these speaker labels need to be associated with words recognized by an ASR system (i.e. ``who spoke what'').
This involves a complicated multi-module architecture with an orchestration algorithm to merge ASR and diarization results based on segment timestamps. Both modules are also required to be synchronized to have similar latency for the best results. To address these challenges, there have been a few pioneering proposals for joint modeling of word-level speaker diarization with ASR, summarized in Section~\ref{sec:related-work}.

Inspired by the multi-output, multi-task learning work~\cite{wang2023multi} and many other ASR-auxiliary joint learning studies~\cite{liu2021improving,zhang2021e2e}, we extend the ASR-auxiliary joint modeling architecture to include an auxiliary network for speaker diarization, with a separate encoder and joint network for predicting speaker labels. We assess our method on various test scenarios: public and simulated data, short-form and long-form audios, 2-speaker and 3+ speaker scenarios, etc. Experiments show that WEEND significantly outperforms the turn-based diarization baseline on Callhome by 25\%, demonstrates superior quality across all 2-speaker short-form test scenarios and generalizes up to 5 minutes of 2-speaker long-form audios with no performance degradation. For speech that involve 3+speakers, WEEND is still capable of predicting speaker labels well provided that it is trained on sufficient in-domain data.


\section{Related work}
\label{sec:related-work}

Shafey et al.~\cite{shafey2019joint} proposed inserting speaker role tags into the transcripts and training like ASR. However, the problem was constrained to 2-speaker doctor-patient conversations. It is solving a word-level doctor-patient classification problem instead of a generic speaker diarization problem. Another related category of work is speaker attributed ASR (SA-ASR), which typically takes the additional input of speaker profiles and identifies speaker profile indices based on an attention mechanism~\cite{kanda2020joint,kanda2021minimum,kanda2021end,kanda2022streaming}. In the absence of enrolled speaker profiles, the SA-ASR model performs speaker clustering on internal embeddings~\cite{kanda2021investigation}. Moreover, SA-ASR involves an inherent turn detection where it segments speech according to speaker change points. In addition, target speaker ASR (TS-ASR)~\cite{zmolikova2017speaker,delcroix2018single,delcroix2019end,kanda2019auxiliary} can also be considered as diarizing target speaker speech via enrolled speaker embedding extraction.

The main contributions and novelty of our work lie in the following aspects: (1) We propose Recurrent Neural Network Transducer (RNN-T) based ASR-diarization multi-task learning,
where both tasks are strongly coupled by sharing blank logits. (2) Our approach leads to a much simpler pipeline, with no requirement of speaker profiles, enrollment or clustering. (3) We can make use of pre-trained ASR systems to quickly adapt to the diarization task, without affecting ASR performance.

\section{System descriptions}
\label{sec:system}

As discussed in Section~\ref{sec:intro}, diarization is particularly helpful when associated with recognized words. Speaker labels are also natively coupled with words. Thus, we model ASR and diarization together. For each utterance, we aim at recognizing the speech words and predicting the corresponding speakers simultaneously. For the target sequence, we tokenize the transcript with a wordpiece model and meanwhile construct a same-length speaker label sequence. Within each utterance, we map raw speaker labels (e.g. ``speaker:A'') to integer-indexed speaker labels in a ``first come, first serve'', order-based fashion. That is, the $N$th speaker that starts speaking is labeled as \texttt{<spk:N>}. In the following sections, we describe the blank sharing multi-output setup in Section~\ref{sec:multi-output}. Section~\ref{sec:proposal} introduces our proposed method and its main modifications.

\subsection{RNN-T with multi-output joint networks}
\label{sec:multi-output}

We follow the RNN-T ASR architecture in Wang et al.~\cite{wang2023multi}.
Specifically, the joint network of an RNN-T model~\cite{graves2012sequence} fuses audio features extracted by the encoder with the text features extracted by the prediction network.
Formally, let the encoder output be $[f_0, \dots, f_{T-1}]$ and the prediction network output be $[g_0, \dots, g_{U-1}]$, where $f_t \in \bbR^{D_a}$, $g_u \in \bbR^{D_l}$. $t$ and $u$ denote the time and label sequence indices, ${D_a}$ and ${D_l}$ denote acoustic and text feature dimensions. The ASR symbol space $\calY$ consists of a special $\blank$ token for non-emission and a set of $V-1$ non-blank wordpieces, i.e., $\calY=\{y^0=\blank, y^1, \dots, y^{V-1}\}$. The joint network hidden embedding $h_{t,u}$ is merged from the acoustic and text features:
\begin{align} \label{eqn:asr-hidden}
    h_{t,u} = P \cdot f_t + Q \cdot g_u + b_h \quad \in \bbR^{D_h}
\end{align}
where $P$, $Q$ are projection matrices and $b_h$ is the bias term. The raw logits $s_{t,u}$ before softmax are computed:
\begin{align} \label{eqn:asr-logits}
    s_{t,u} = A \cdot \text{tanh} (h_{t,u}) + b_s \quad \in \bbR^V
\end{align}
where $A$ is the projection matrix, $b_s$ is the bias term. We use the hybrid auto-regressive transducer model~\cite{variani2020hybrid} and the factorized posterior probability distribution over $\calY$ can be formulated as:
\begin{align} \label{eqn:asr-posterior}
    P_{t,u}(y^v|f_{0:t}, g_{0:u}) 
    & = (1 - b_{t,u}) \cdot  \text{softmax} (s_{t,u}[1\!:])[v-1]
\end{align}
for $v=1,\dots, V-1$, with $y^v \neq \blank$, where the factorized blank distribution $b_{t,u}$ is defined as:
\begin{align} \label{eqn:prob-blank}
    b_{t,u} := P_{t,u}(\blank|f_{0:t}, g_{0:u}) = \text{sigmoid}(s_{t,u}[0])
\end{align}




To extend the RNN-T architecture for auxiliary tasks, we introduce additional last linear layer parameters $A^{\text{aux}}$ and $b_s^{{\text{aux}}}$.
Blank logits are shared between ASR and the auxiliary task to ensure output synchronization across tasks. Denote the auxiliary task output space is $\calY_{\text{aux}}= \{\blank, y_\text{aux}^1, \dots, y_\text{aux}^{V_\text{aux}-1}\}$, where $V_\text{aux}$ is the size of the auxiliary label space including the shared blank. We re-use the blank logits from ASR~\eqref{eqn:asr-logits} and the auxiliary task raw logits can be expressed this way:
\begin{align}
    s_{t,u}^{\text{aux}} = [s_{t,u}[0],\ \  A^{\text{aux}} \cdot \text{tanh} (h_{t,u}) + b_s^{{\text{aux}}}] \quad \in \bbR^{V_{\text{aux}}}.
\end{align}

For the inference procedure, ASR blank emissions are directly shared. Whenever the ASR beam search emits a non-blank, we apply softmax on the auxiliary logits $s_{t,u}^{\text{aux}}$ to obtain the probabilities over the auxiliary label space $\calY_{\text{aux}}$:
\begin{gather} \label{eqn:aux-softmax}
    P_{t,u} (\calY_{\text{aux}}|\text{non-blank}, f_{0:t}, g_{0:u}) = \text{softmax}(s_{t,u}^{\text{aux}}[1\text{:}])
\end{gather}
and select the auxiliary label with $\text{argmax}(\cdot)$ on the softmax logits.

\subsection{Proposed joint ASR and diarization system}
\label{sec:proposal}

\begin{figure}
	\centering
    \includegraphics[trim={0.1in 0in 0in 0in},height=145pt]{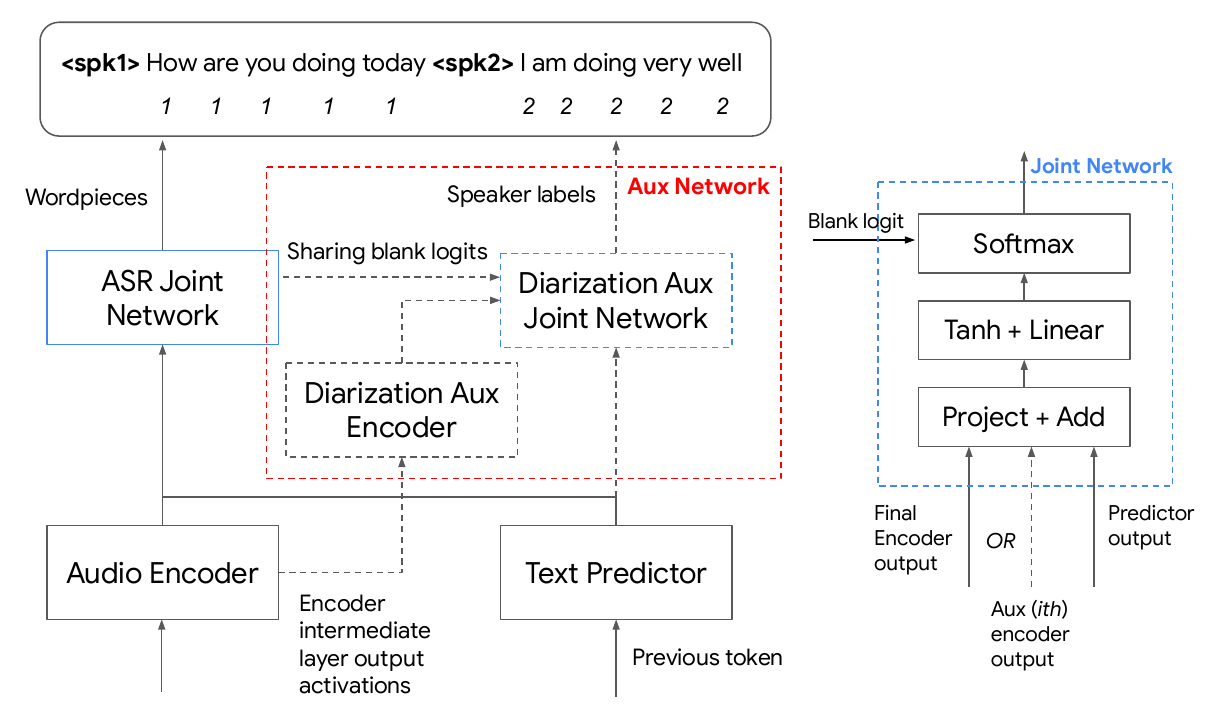}
    \label{fig:diagram1}
    \vspace{-3mm}
    \caption{Diagram of the proposed ASR-diarization joint architecture with an auxiliary network of an auxiliary encoder and a separate joint network sharing ASR blank logits.
    Dotted lines indicate the newly added designs. Solid outlined modules are frozen during training.}
    
    \label{fig:diagram}
\end{figure}


In the context of speaker diarization, auxiliary labels are now speakers. The synchronization between ASR and diarization perfectly enables the word-level speaker diarization task. Therefore, we further extend the multi-output joint architecture from Section~\ref{sec:multi-output}, and propose this neural architecture with the addition of an auxiliary network in Figure~\ref{fig:diagram} including an auxiliary diarization encoder, intermediate layer activations wiring and an auxiliary joint network. The necessity of these is further discussed in Section~\ref{sec:ablation}.

\vspace{-1mm}
\subsubsection{Auxiliary diarization encoder}
\label{sec:aux-encoder}
\vspace{-1mm}
We insert a second, auxiliary encoder for diarization, which generates a new encoder feature output $f^{\text{aux}}_t \in \bbR^{D_{\text{aux}}}$ via Eq.~\eqref{eqn:asr-hidden}. The motivation of this encoder is to extract speaker-related features. The encoder can be any model architecture capable of modeling temporal context, such as LSTM~\cite{hochreiter1997long}, transformer or conformer.

\vspace{-1mm}
\subsubsection{Intermediate activations}
\vspace{-1mm}
We feed the ASR encoder intermediate activations as the input to the auxiliary diarization encoder. According to these work~\cite{zhang2021e2e,zhang2022intermediate}, we expect the model to perform the best when using intermediate layer outputs instead of raw audio features or last layer outputs.

\vspace{-1mm}
\subsubsection{Auxiliary joint network}
\vspace{-1mm}
We define a separate auxiliary joint network for diarization label sequence prediction. Following the notation~\eqref{eqn:asr-hidden}, we introduce the auxiliary parameters $P_\text{aux}$, $Q_\text{aux}$ and $b_{h\text{aux}}$ to learn the hidden embedding from the acoustic and text features, as well as the final projection parameters $A^{\text{aux}}$ and $b_s^{{\text{aux}}}$. 


\section{Experiments}
\label{sec:exp}

\subsection{Datasets}

\begin{table}
    \centering
    \caption{Diarization public and simulated datasets statistics}
    \vspace{-1mm}
    \begin{adjustbox}{max width=\linewidth}
    \begin{tabular}{ccccccc}
    \toprule
    \multirow{2}{*}{Datasets} & \multirow{2}{*}{Domain} & \multirow{2}{*}{\# Spk} & \multicolumn{2}{c}{Avg length (sec)} & \multicolumn{2}{c}{Total hours (hr)} \\
     & & & Train & Eval & Train & Eval\\
    \midrule
    AMI & Meeting & 3-4 & 15/30/60 & 2039 & 81 & 9.1 \\
    Callhome & Telephone & 2 & 15/30/60 & 301 & 14 & 1.7 \\
    Fisher & Telephone & 2 & 15/30/60 & 600 & 1920 & 28.7 \\
    Sim 2spk & Read & 2 & 57.9 & 36.1 & 6434 & 39.7 \\
    Sim 3spk & Read & 3 & 68.5 & 43.1 & 7137 & 43.3 \\
    Sim 4spk & Read & 4 & 94.2 & 59.4 & 9848 & 57.0 \\
    \bottomrule
    \end{tabular}
    \end{adjustbox}
    \vspace{-3mm}
    \label{table:datasets}
\end{table}

Our public datasets are listed in the first 3 rows of Table~\ref{table:datasets}:
\begin{enumerate}[leftmargin=*]
    \item AMI~\cite{carletta2005ami}: we use the official ``Full-corpus-ASR partition'' training and test subsets. The speaker label ground truth was obtained based on the word-level annotations in the v1.6.2 AMI manual annotation release.
    \item Callhome~\cite{canavan1997callhome}: we use the Callhome American English Speech official training and evaluation subsets.
    \item Fisher~\cite{cieri2004fisher}: we withhold a test subset of 172 utterances\footnoteurl{https://github.com/google/speaker-id/blob/master/publications/Multi-stage/evaluation/Fisher/eval_whitelist.txt} from Fisher English Training Speech and use the rest for training.
\end{enumerate}

The training data are segmented into 15, 30 and 60 second segments for model training. The segmentation avoids chopping in the middle of a sentence. For the test splits, we evaluate not only the original full-length audios but also their short-form segments of 30, 60 and 120 seconds. The official metadata RTTMs are converted to target word and speaker label sequences.

Besides public data, we simulate multi-speaker utterances from LibriSpeech for data augmentation to mitigate the lack of training data. For each simulated conversation, we sample $M$ unique speakers and $N$ utterances from each speaker, and randomly drop $K=0,1,2$ utterances for variety. Remaining samples are concatenated in random order, with inserted pause (uniform from 0.2 to 1.5 seconds) and cross-fade (uniform from 0 to 0.2 seconds). There are both real and fake speaker turns from this simulation. Simulated data statistics are also listed in Table~\ref{table:datasets}. For training sets, we sample from LibriSpeech train-clean-100h, train-clean-360h and train-other-500h. For test sets, we sample from test-clean and test-other.


\subsection{Model architecture}
\label{sec:config}

We extract 128-dimensional log-Mel spectrogram features using a 32ms Hamming window with a hop length of 10ms. Every 4 consecutive frames are stacked and sub-sampled by a factor of 3 to generate 512-dimensional features at 30ms frame rate. The input features are first fed to the causal ASR encoder, which consists of 12 conformer layers~\cite{gulati2020conformer} of 512 dimensions, with 8-head attention, convolutional kernel size of 15 and a left context of 23 frames. Funnel pooling~\cite{dai2020funnel}, with a downsampling factor of 2, is placed after the fourth conformer layer. The ASR encoder outputs $D_a=512$ feature dimensions. The decoder is an embedding-based prediction network which computes language model features of $D_l=640$ dimensions, based on two previous non-blank tokens~\cite{Rami21}. The ASR joint network hidden dimension is $D_h=640$. A last linear layer projects the hidden dimension to the wordpiece model vocabulary size $V=4096$.

As for the auxiliary network, the 5th ASR Conformer layer outputs are fed to our LSTM encoder, which has a stacked of 9 layers, each layer containing 1024 hidden nodes and 512 output nodes. The auxiliary joint is set to have a hidden dimension of $D^{\text{aux}}_h=640$, and the same prediction network outputs are used in the auxiliary joint. We pre-define a speaker set from $1$ to $N=8$, so the last linear layer projects the joint hidden embedding to $V_\text{aux}-1=8$.
For training, we initialize the ASR audio encoder, text predictor, and joint network from this pre-trained ASR model~\cite{Rami21} and freeze the parameters of these components (i.e. only the auxiliary network parameters are updated during training). The loss function is a standard RNN-T loss but we only include the speaker label RNN-T loss (with the shared blank logits), because the ASR RNN-T loss acts more like a scaling factor when ASR is frozen.


\subsection{Baseline: turn-based diarization}
\label{sec:baseline}
We set up the turn-based diarization baseline following the ``turn-to-diarize'' system~\cite{xia2022turn,zhao2023scd} without pairwise speaker turn constraints and apply multi-stage clustering~\cite{wang2022highly}. We pair it with the same RNN-T ASR model described in Section~\ref{sec:config} to retrieve word-level speaker labels, specifically by assigning speakers to the recognized words based on the maximum speaker overlap in duration for each word.


\subsection{Metrics}
\label{sec:metrics}
We report the Word Error Rate (WER) for ASR quality, and the Word Diarization Error Rate (WDER) from~\cite{shafey2019joint} for diarization quality. This WDER is a time-invariant diarization error metric that does not take time boundaries into consideration, and it suits the word-level end-to-end diarization problem better.
In addition, datasets like AMI have lots of overlaps but the ASR system by itself does not support overlapping speech. This leads to considerable label confusion around overlapping speech and quick speaker changes. Hence, for AMI in this paper, we evaluate and report a modified WDER\footnote{The modified metric algorithm and unmodified AMI metrics are reported at \url{https://github.com/google/speaker-id/blob/master/publications/WEEND/README.md}} which does not count words that overlap with any other word in the ground truth. We gather statistics of how many words are dropped to ensure there are still enough words left from each utterance. In our evaluation, the word drop percentage distribution on AMI has a mean of 10\% and standard deviation of 14\%.


\subsection{Experimental results}
\label{sec:results}

\begin{table}
  \centering
  \caption{ASR and diarization performance of the baseline and proposed models. We report WERs (\%) followed by the detailed breakdowns of substitution (S), deletion (D) and insertion (I) error rates.}
        \begin{tabular}{c|c|cc}
           \multirow{2}{*}{Testsets} & \multirow{2}{*}{WER (S/D/I)} & \multicolumn{2}{c}{WDER (\%)} \\
            & & Baseline & Proposed \\ \midrule
           Callhome         & 45.9 (12.8/9.7/23.3) & 10.3 & 7.7 \\
           Fisher           & 20.5 (8.7/10.4/1.4) & 3.6 & 8.0 \\
           AMI              & 29.6 (8.9/19.9/0.8) & 8.7 & 50.0 \\
           Sim 2spk         & 8.1 (6.4/1.0/0.7) & 4.2 & 4.1 \\
           Sim 3spk         & 8.3 (6.5/1.0/0.8) & 4.2 & 3.6 \\ 
           Sim 4spk         & 8.1 (6.4/1.0/0.7) & 4.5 & 5.1 \\
        \end{tabular}
  \vspace{-3mm}
  \label{table:proposed-result}
\end{table}

\begin{table}
  \centering
  \caption{Short-form test WDER (\%) on various audio duration.}
    \vspace{-1mm}
        \begin{tabular}{c|c|cc}
          \multirow{2}{*}{Testsets} & Short-form & \multicolumn{2}{c}{WDER (\%)} \\
            & Lengths (s) & Baseline & Proposed \\ \midrule
            \multirow{3}{*}{Callhome} & 30 & 13.6 & 9.3 \\
                                      & 60 & 9.8 & 8.9 \\
                                      & 120 & 10.5 & 8.9 \\ \midrule
            \multirow{3}{*}{Fisher}   & 30 & 8.6 & 3.8 \\
                                      & 60 & 4.8 & 3.7 \\
                                      & 120 & 4.0 & 3.7 \\ \midrule
            \multirow{3}{*}{AMI}      & 30 & 10.1 & 9.9 \\
                                      & 60 & 6.7 & 13.3 \\
                                      & 120 & 8.0 & 18.8 \\ 
        \end{tabular}
  \label{table:short-form}
\end{table}




The ASR and diarization quality is reported in Table~\ref{table:proposed-result}. For ASR, Callhome has high insertion error because its ground truth transcript contains many missing words. The deletion error rate is high due to the fact that ASR does not support overlapping speech. The annotation standard of the diarization datasets (e.g. ``mhm'', ``y- ye- yes'') adds to the deletions and substitutions. On simulated datasets, the substitution rate is high because ASR is not trained on read speech like LibriSpeech. For diarization, our model outperforms the baseline on 5-min Callhome test data by 25\%. The performance degrades on longer audios such as 10-min Fisher test utterances. AMI has the hardest test cases: longest audio, more speakers, overlapping speech. We investigate each of those aspects in-depth.

\subsubsection{Short-form and audio duration}
\label{sec:audio-duration}

Table~\ref{table:short-form} presents the short-form performance. Our proposed model outperforms the baseline on short-form Callhome and Fisher significantly, particularly on 30s segments by 32\% and 56\%. However, the limitation is that the performance degrades if the audio is very long (e.g. Fisher), and it is most notable on AMI test data, which is the longest. This limitation stems from the mismatch between training segments and the full length test segments. In contrast, the baseline does not suffer from this issue because it is an unsupervised, clustering based model. In fact, the baseline quality is sub-optimal when the audio is only 30 seconds. This is because each segment can be 6 seconds, and a 30-second utterance might end up with only 5 segments, which is not enough for clustering to achieve good results.

\subsubsection{Variable number of speakers}

To understand the effect of number of speakers, we break down the composition of AMI WDER bucketed by the number of speakers. Table~\ref{table:ami-num-spks} shows that our model performs better than the baseline on 1 or 2 speaker cases (except for the 120-sec testset). The majority of errors come from 3 or 4 speaker cases. Since Table~\ref{table:proposed-result} proves that the model is indeed capable of learning 3+speaker diarization on simulated testsets, we believe the poor performance comes from (1) very limited amount of in-domain training data (only 80 hours) (2) quick speaker changes in meeting conversations.
The second point matches our loss analysis observations: if there is no interruption or background speaker, speaker labels are predicted correctly, consistently. But speaker predictions have much higher chances to be wrong around quick speaker change points. This is likely because ASR in our proposed system only emits a single speaker speech on overlapping speech. It confuses model training with which speaker label to predict on nearby words.




\begin{table}
  \centering
  \caption{Pre-segmented short-form AMI WDER (\%) of the baseline and proposed models, breakdown by reference number of speakers. For each testset, we compute the WDER for each group of utterances with the same number of ground truth speakers. For the evaluation on 120-sec segments, since there are only 6 speaker test examples, we do not list these results.}
        \begin{tabular}{c|cccc|cccc}
            AMI & \multicolumn{4}{c|}{Baseline WDER (\%)} & \multicolumn{4}{c}{Proposed WDER (\%)} \\
             Lengths & \!1spk\!\! & \!\!2spk\!\! & \!\!3spk\!\! & \!\!4spk\!\! & \!1spk\!\! & \!\!2spk\!\! & \!\!3spk\!\! & \!\!4spk\!\! \\ \midrule
           30-sec & 18.6 & 10.0 & 8.8 & 8.4 & 1.1 & 5.8 & 10.1 & 15.5 \\
           60-sec & 10.8 & 6.3 & 5.6 & 6.9 & 0.8 & 5.2 & 12.2 & 17.1 \\
           120-sec & - & 6.4 & 4.4 & 9.3 & - & 9.8 & 15.8 & 20.8 \\
        \end{tabular}
  \label{table:ami-num-spks}
\end{table}

\subsection{Ablation studies}
\label{sec:ablation}

\begin{table}
  \centering
  \caption{Impact of intermediate layer selection on WDER (\%). Callhome is abbreviated as CH. Average numbers are reported for simulated and short-form AMI.}
        \begin{tabular}{l|cccc}
            Intermediate & \multirow{2}{*}{CH} & \multirow{2}{*}{Fisher} & \multirow{2}{*}{Sim} & AMI  \\
            Layer Selection & & & & Short \\ \midrule
            0th Conf layer (features) & 23.8 & 24.5 & 10.4 & 22.8 \\
            5th Conf layer (proposed) & 7.7 & 8.0 & 4.3 & 14.0 \\
            12th Conf layer (last) & 33.6 & 37.3 & 46.9 & 27.5 \\
        \end{tabular}
  \label{table:ablation-model}
\end{table}

%

We explore how much the network architecture affects model performance. If we completely remove the auxiliary encoder (i.e. use the ASR encoder output along with a separate joint network directly), the model simply does not learn diarization properly. If we keep the auxiliary encoder, intermediate layer output selection leads to different outcomes. As shown in Table~\ref{table:ablation-model}, neither the first layer nor the last layer works well. This aligns with our expectation and conclusions from previous work~\cite{zhang2021e2e,zhang2022intermediate}. ASR encoder tends to discard speaker knowledge towards the last layer. Raw features are also hard for training to converge to the optimal space.
There exists a sweet spot where a certain intermediate layer works the best.

\begin{table}
  \centering
  \caption{Training data augmentation impact on WDER (\%). The second row excludes simulated data from training. The last row further drops 30/60s training segments, i.e. only trained on 15s data.}
        \begin{tabular}{l|cccccc}
            \multirow{2}{*}{Model}
             & \multirow{2}{*}{\!CH\!} & \!CH\! & \multirow{2}{*}{\!Fisher\!} & \!Fisher\! & \multirow{2}{*}{\!Sim\!} & \!AMI\! \\
             & & \!Short\! & & \!Short\! & & \!Short\! \\ \midrule
            Proposed & 7.7 & 9.0 & 8.0 & 3.7 & 4.3 & 14.0 \\
            {} {} -Simulated & 11.6 & 9.8 & 12.3 & 5.4 & 22.2 & 19.0 \\
            {} {} {} {} -30/60s segs & 28.8 & 22.5 & 22.1 & 15.7 & 26.8 & 26.2 \\
        \end{tabular}
  \label{table:ablation-data}
\end{table}

Table~\ref{table:ablation-data} summarizes our data ablation studies.
Even though the simulated data is from a different domain, it still effectively mitigates the public data insufficiency. Fisher and Callhome short-form improvements are the least, most likely due to the fact that we have enough 2-speaker telephony training data from Fisher (almost 2k hours). Furthermore, if the training data only include 15s segments, the model quality degrades dramatically on all testsets. This matches the audio duration generalization discussion in Section~\ref{sec:audio-duration} and suggests that we should include longer training segments if feasible.


\section{Conclusions}
\label{sec:conclusions}

This paper proposed and studied word-level end-to-end neural diarization via auxiliary networks, without involving turn-based segmentation, speaker profiles or clustering. This approach presents promising opportunities over conventional unsupervised methods, where we observed superior performance on 2-speaker short-form scenarios. The limitation is that due to the lack of 3+speaker, long-form in-domain training data, the model does not generalize well enough to those cases.
In the future, we would like to investigate chunk-aware learning for long-form training with historical context for long-form generalization. Additionally it is essential to develop advanced data augmentation techniques to simulate large amounts of in and out of domain conditions with arbitrary number of speakers. 
Lastly, this architecture can be further extended to be support overlapping speech with serialized output training (SOT).

\clearpage
\bibliographystyle{IEEEbib}
\bibliography{refs}


\end{document}